\documentclass{article}
\usepackage{dcase2019,amsmath,graphicx,url,times,booktabs,multirow,tabularx,amssymb,multirow}
\usepackage{subcaption}

\title{IMPROVING SOUND EVENT DETECTION IN DOMESTIC ENVIRONMENTS USING SOUND SEPARATION}
\name{
    Nicolas Turpault$^1$\thanks{Part of this work was made with the support of the French National Research Agency, in the framework of the  project LEAUDS “Learning to under-stand audio scenes” (ANR-18-CE23-0020) and the French region Grand-Est. Experiments presented in this paper were carried out using the Grid5000 testbed, supported by a scientific interest group hosted by Inria and including CNRS, RENATER and several Universities as well as other organizations (see https://www.grid5000).},
    Scott Wisdom$^2$,
    Hakan Erdogan$^2$,
    John R. Hershey$^2$,
    }
\secondlinename{
Romain Serizel$^1$,
      Eduardo Fonseca$^3$,
    Prem Seetharaman$^4$,
    Justin Salamon$^5$
    }
\address{$^1$ Universit{\'e} de Lorraine, CNRS, Inria, Loria, F-54000 Nancy, France \\
        $^2$ Google Research, AI Perception, Cambridge, United States\\
        $^3$ Music Technology Group, Universitat Pompeu Fabra, Barcelona\\
        $^4$ Interactive Audio Lab, Northwestern University, Evanston, United States\\
        $^5$ Adobe Research, San Francisco, United States
 }

\begin{document}

\ninept
\maketitle

\begin{sloppy}

\begin{abstract}
Performing sound event detection on real-world recordings often implies dealing with overlapping target sound events and non-target sounds, also referred to as interference or noise. Until now these problems were mainly tackled at the classifier level. We propose to use sound separation as a pre-processing for sound event detection. In this paper we start from a sound separation model trained on the Free Universal Sound Separation dataset and the DCASE 2020 task 4 sound event detection baseline. We explore different methods to combine separated sound sources and the original mixture within the sound event detection. Furthermore, we investigate the impact of adapting the sound separation model to the sound event detection data on both the sound separation and the sound event detection.

\end{abstract}

\begin{keywords}
Sound event detection, synthetic soundscapes, sound separation
\end{keywords}

\section{Introduction}
\label{sec:intro}
Sound event detection (SED) is the task of describing, from an audio recording, what happens and when each single sound event is occurring~\cite{virtanen2018computational}. This is something that we, as humans, do rather naturally to obtain information about what is happening around us. However, trying to reproduce this with a machine is not trivial, as the SED algorithm needs to cope with several problems, including audio signal degradation due to additive noise or overlapping events~\cite{benetos_detection_2016}. Indeed, in real-world scenarios, the recordings provided to the SED systems contain not only target sound events, but also sound events that can be considered as ``noise'' or ``interference.'' Also, several target sound events can occur simultaneously.

In the past, the overlapping sound events problem has been tackled from the classifier point of view. This can be done by training the SED as a multilabel system in which case the most energetic sound events are usually detected more accurately than the rest~\cite{salamon2015feature,serizel_2020}. Some other approaches tried to deal more explicitly with this problem using either a set of binary classifiers~\cite{mesaros2016tut}, using factorization techniques on the input of the classifier~\cite{benetos2016detection,bisot2017overlapping}, or exploiting spatial information when available~\cite{adavanne2018multichannel}. The additive noise problem is usually solved by training SED systems on noisy signals. This may be effective to some degree when the noise level is low, but much less so when the noise level increases~\cite{serizel_2020}.

Sound separation (SS) seems like a natural candidate to solve these two issues.
SS systems are trained to predict the constituent sources directly from mixtures. Thus, sound separation can both decrease the level of interfering noise and enable a SED system to detect quieter events in overlapping acoustic mixtures.
Until recently, SS has been mainly applied to specific classes of signals, such as speech or music. However, recent works has shown that sound separation can also be applied to separating sounds of arbitrary classes, a task known as ``universal sound separation''~\cite{kavalerov2019universal,tzinis2020improving,olvera:hal-02567542}.

In this paper, we propose to combine a universal SS algorithm~\cite{kavalerov2019universal,tzinis2020improving} used as a pre-processing to the DCASE 2020 SED baseline~\cite{turpault2020}. We investigate the impact of the data used to train the SS on the SED performance. We also explore different ways to combine the separated sound sources at different stages of SED.

\section{Problem and baselines description}
\label{sec:pb}

\begin{figure*}
    \centering
\begin{subfigure}[b]{0.16\textwidth}
  \centering
  \includegraphics[width=\columnwidth]{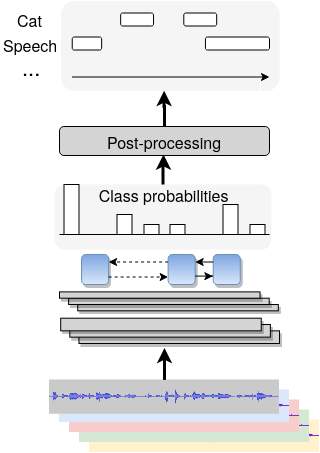}
  \caption{Early integration.}
\label{fig:sed_early}
\end{subfigure}\hfill
    \begin{subfigure}[b]{0.35\textwidth}
\centering
  \includegraphics[width=\columnwidth]{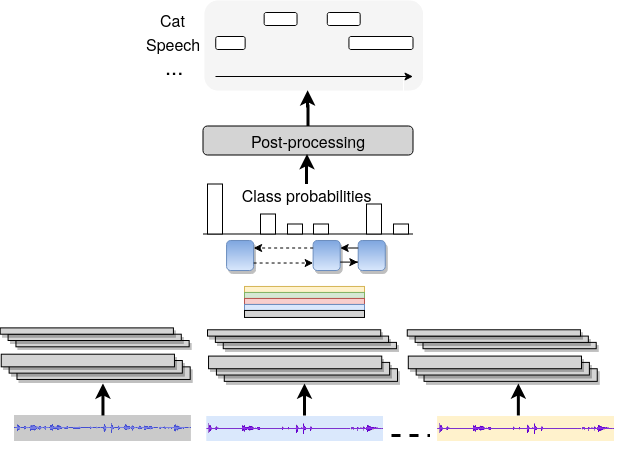}
  \caption{Middle integration.}
\label{fig:sed_middle}
\end{subfigure}\hfill
    \begin{subfigure}[b]{0.41\textwidth}
\centering
  \includegraphics[width=\columnwidth]{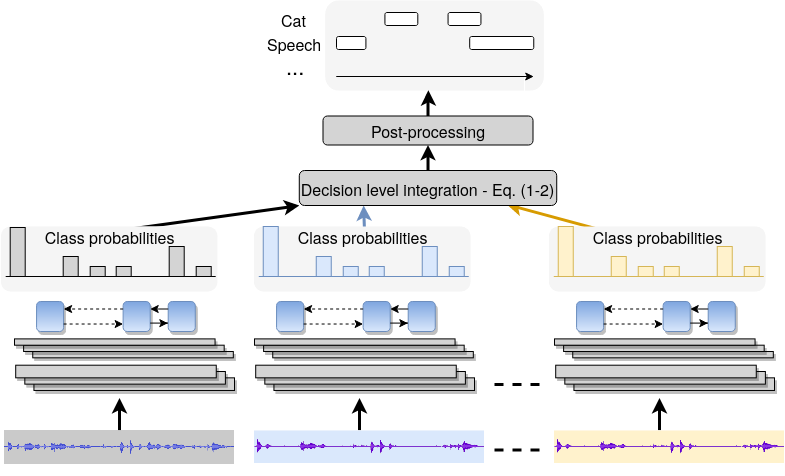}
  \caption{Late integration.}
\label{fig:sed_late}
\end{subfigure}\hfill
\caption{Integration between the SS and SED (gray waveform represents mixture, and colored waveforms represent separated sources).}
\end{figure*}

% \subsection{Problem description}
We aim to solve a problem similar to that of DCASE 2019 Task 4~\cite{turpault_2019}. Systems are expected to produce strongly-labeled outputs (i.e.~detect sound events with a start time, end time, and sound class label), but are provided with weakly labeled data (i.e.~sound recordings with only the presence/absence of a sound event included in the labels without any timing information) for training. Multiple events can be present in each audio recording, including overlapping target sound events and potentially non-target sound events. Previous studies have shown that the presence of additional sound events can drastically decrease the SED performance~\cite{serizel_2020}.

\subsection{Sound event detection baseline}
\label{sub:sed_baseline}
The SED baseline system uses a mean-teacher model which is a combination of two models: a student model and a teacher model (both have the same architecture). The student model is the final model used at inference time, while the teacher model is aimed at helping the student model during training and its weights are an exponential moving average of the student model's weights. A more detailed description can be found in Turpault and Serizel~\cite{turpault2020}.

\subsection{Sound separation baseline}
% \begin{figure}
% \centering
%   \includegraphics[width=0.6\columnwidth]{sound_sep.png}
%   \caption{Sound source separation.}
% \label{fig:ss_prepro}
% \end{figure}
The baseline SS model uses a similar architecture to an existing approach for universal sound separation with a fixed number of sources \cite{kavalerov2019universal, tzinis2020improving}, which employs a convolutional masking network using STFT and analysis and synthesis.
%Additionally, we use a weighted mixture consistency layer \cite{wisdom2019differentiable}, where the weights are predicted by the network, with one scalar per source, that ensures the separated sources add up to the original mixture.
The training loss is negative stabilized signal-to-noise ratio (SNR) \cite{wisdom2020unsupervised} with a soft-threshold $\text{SNR}_{\text{max}}$.
Going beyond previous work, the model in this paper is able to handle variable number sources by using different loss functions for active and inactive reference sources that encourage the model to only output as many nonzero sources as exist in the mixture. Additional source slots are encouraged to be all-zero.

\section{Sound event detection and separation}

\label{sec:sed_ss}
\subsection{Sound separation for sound event detection}
Overlapping sound events are typically more difficult to detect as compared to isolated ones.
SS can be used for SED by first separating the component sounds in a mixed signal and then applying SED on each of the separated tracks. The decisions obtained on separated signals may be more accurate than the ones on the mixed signal. On the other hand, separation of sounds is not a trivial problem and may introduce artifacts which in turn may make sound SED harder. So, it is necessary to jointly investigate SS and SED.

\subsection{Sound event detection on separated sources}

In the approaches described here, SS provides several audio clips that contain information related to the sound sources composing the original (mixture) clip. Each of these new audio clips (separated sound sources) are used together with the mixture clip within the SED. We compare three different approaches to integrate the information from these audio clips at different levels of the model.%: Early integration, embedding level integration and decision level integration.

\subsubsection{Early integration}
This approach is similar to the SED baseline except that all the audio clips (mixture and separated sound sources) are concatenated as input channels to form a new tensor (Figure~\ref{fig:sed_early}). The first channel always contains the mixture clip while the  separated sound source clips are provided with no particular order. The model is trained like the SED baseline using the annotations of the mixture clip. %At test time, the model is provided with the mixture clip and the separated sound source clips concatenated as input channels to the CRNN.
% \begin{figure}
% \centering
%   \includegraphics[width=0.5\columnwidth]{early.png}
%   \caption{Early integration.}
% \label{fig:sed_early}
% \end{figure}
\subsubsection{Middle integration}
We re-use the CNN block from the SED baseline to extract embeddings from the mixture clip and the separated sound sources clips (Figure~\ref{fig:sed_middle}). The embeddings are concatenated along the feature axis and fed into a fully connected layer before training a new RNN classifier within a mean-teacher student framework. %The classifier is used at test time to perform the SED.
% \begin{figure}
% \centering
%   \includegraphics[width=\columnwidth]{middle.png}
%   \caption{Embedding level integration.}
% \label{fig:sed_middle}
% \end{figure}
\subsubsection{Late integration}
For this approach, we apply the SED baseline on the mixture clip and the separated sound source clips (Figure~\ref{fig:sed_late}). The SED output for each of these clips are obtained from the $y_{\mathrm{DSL},c}$, the raw outputs of the classifier corresponding to each sound class $c$ among the $C$ sound classes. The combined raw output (before thresholding and post-processing) for each class $c$ is obtained as follows:
\begin{equation}
y_{\mathrm{DSL},c} = \big(\frac{y_{M,c}^{q} + y_{SS,c}^{q}}{2}\big)^{1 / q}
\end{equation}
where $y_{M,c}$ and $y_{SS,c}$ are the raw classifier outputs for the sound class $c$ obtained on the mixture clips and the separated sound source clips, respectively. The sound source/mixture combination weight is $q$. The classifier output for the sound class $c$ is obtained from the raw classifier outputs on each individual separated sound sources as follows:
\begin{equation}
y_{SS,c} = \big(\frac{1}{N_s} \sum_{s=1}^{N_s} \mathbf{y}_{s,c}^{p}\big)^{1 / p}
\end{equation}
where $N_s$ is the number of separated sound source clips obtained from the SS, $y_{s,c}$ is the raw classifier output for the sound class $c$ obtained for the separated sound source clip $s$ and $p$ is the sound sources combination weight.
% \begin{figure}
% \centering
%   \includegraphics[width=\columnwidth]{late.png}
%   \caption{Decision level integration.}
% \label{fig:sed_late}
% \end{figure}
\section{Baselines setup and dataset}
\label{sec:base_bdd}
\subsection{DESED dataset}
The dataset used for the SED experiments is DESED\footnote{\url{https://project.inria.fr/desed/}}, a flexible dataset for SED in domestic environments composed of 10-sec audio clips that are recorded or synthesized~\cite{turpault_2019,serizel_2020}. The recorded soundscapes are taken from AudioSet~\cite{Gemmeke2017audioset}. The synthetic soundscapes are generated using Scaper~\cite{salamon2017scaper}. The foreground events are obtained from FSD50k~\cite{font2013freesound,fonseca2020fsd50k}. The background textures are obtained from the SINS dataset (activity class ``other'')~\cite{Dekkers2017} and TUT scenes 2016 development dataset~\cite{mesaros_tut_2016}.

The dataset includes a synthetic validation set simulated from different isolated those in the training set (SYN\_VAL) , a validation set  and a public evaluation set composed of recorded clips (REC\_VAL and REC\_EVAL) that are used to adjust the hyper-parameters and evaluate the SED, respectively.

\subsection{FUSS dataset}

The Free Universal Sound Separation (FUSS)\footnote{\url{https://github.com/google-research/sound-separation/tree/master/datasets/fuss}} dataset \cite{wisdom2020fuss} is intended for experimenting with universal sound separation \cite{kavalerov2019universal}, and is used as training data for the SS system.
Audio data is sourced from \url{freesound.org}. Using labels from FSD50k \cite{fonseca2020fsd50k}, gathered through the Freesound Annotator \cite{Fonseca2017freesound}, these source files have been screened such that they likely only contain a single type of sound. Labels are not provided for these source files, and thus the goal is to separate sources without using class information.
%and are not considered part of the challenge. For the purpose of the DCASE Task4 Sound Separation and Event Detection challenge, systems should not use FSD50K labels, even though they may become available upon FSD50K release.
To create reverberant mixtures, 10 second clips of sources are convolved with simulated room impulse responses. Each 10 second mixture contains between 1 and 4 sources. Source files longer than 10 seconds are considered "background" sources. Every mixture contains one background source, which is active for the entire duration.

\subsection{Sound event detection baseline}
The SED baseline\footnote{\url{https://github.com/turpaultn/dcase20_task4/}} architecture and parameters are described extensively in Turpault et al.~\cite{turpault2020}. The performance obtained with this baseline on DESED is presented in Table~\ref{tab:sed_base}.

\begin{table}[]
    \centering
    \caption{Performance for the SED baseline~\cite{turpault2020} on DESED.}
\begin{tabular}{|l|c|c|}
\hline & F1-Score & PSDS \\
  \hline
REC\_VAL & 37.8&	0.540\\
REC\_EVAL&39.0 &0.552\\\hline
SYN\_VAL &62.6&0.695\\
%Synthetic Eval &48.9&0.532\\
\hline
\end{tabular}
\label{tab:sed_base}
\end{table}
% input features for the SED baseline are 128 mel-spectrograms. The mel-spectrogram feature are obtained from Short-term Fourier transform coefficient (STFT) computed on 2048 sample windows with 255 hop size.

% The CNN block is composed of 7 layers with [16,  32,  64,  128,  128, 128, 128] filters per layer, respectively. We use a kernel size of 3x3 and the max-pooling is [[2, 2], [2, 2], [1, 2], [1, 2], [1, 2], [1, 2], [1, 2]] per layer, respectively. The convolution operations are followed by gated liner unit activation.

% The RNN block is composed of 2 layers of 128 bidirectional gated recurrent units. The RNN block is followed by an attention layer that is multiplication between a linear layer with softmax activation and linear layer with sigmoid activation.

% The model is trained with Adam optimizer, we apply 50~\% dropout and early stopping with 10 epochs patience.

% When they are used, the SED detection thresholds are fixed to 0.5 for every classes. The post-processing is a median filtering on $\approx$0.45~s (27 frames at 16~kHz).

\subsection{Sound separation baseline}
The SS system is trained on 16-kHz audio\footnote{ \url{https://github.com/google-research/sound-separation/tree/master/models/dcase2020_fuss_baseline}}. The input to the SS network is the magnitude of the STFT using window size 32ms and hop of 8ms. These magnitudes are processed by an improved time-domain convolutional network (TDCN++) \cite{kavalerov2019universal, tzinis2020improving}, which is similar to Conv-TasNet \cite{luo2019conv}. Like Conv-TasNet, the TDCN++ consists of four repeats of 8 residual dilated convolution blocks, where within each repeat the dilation of block $\ell$ is $2^\ell$ for $\ell=0,..,7$. The main differences between the TDCN++ and Conv-Tasnet are (1) bin-wise normalization instead of global layer normalization, which averages only over basis frames instead of frames and frequency bins, (2) trainable scalar scale parameters multiplied after each dense layer, which are initialized with $0.9^i$, and (3) additional residual connections between blocks, with connection pattern $0\rightarrow 8$, $0\rightarrow 16$, $0\rightarrow 24$, $8\rightarrow 16$, $8\rightarrow 24$, $16\rightarrow 24$.

This TDCN++ network predicts four masks that are the same shape as the input STFT. Each mask is multiplied with the complex input STFT, and a source waveform is computed by applying the inverse STFT. A weighted mixture consistency projection layer \cite{wisdom2019differentiable} is applied to the separated waveforms
%$\underbar{\bf s}_{1:4}$
to be consistent with the input mixture waveform
%${\bf x}$,
where the per-source weights
%$w_j$
are predicted by an additional dense layer using the penultimate output of TDCN++. %network.
% \begin{equation}
%     \hat{s}_j = \underbar{\bf s}_j + \frac{w_j}{\sum_{j'} w_{j'}}
%     \left(
%     {\bf x} - \sum_{j'} \underbar{\bf s}_j
%     \right)
% \end{equation}

To separate mixtures with variable numbers of sources, different loss functions are used for active and inactive reference sources. For active reference sources (i.e. non-zero reference source signals), the soft-threshold for SNR is 30 dB, equivalent to the error power being below the reference power by 30 dB. For non-active reference sources (i.e. all-zero reference source signals), the soft-threshold is 20 dB measured relative to the mixture power, thus gradients are clipped when the error power is 20 dB below the mixture power. Thus, for a $N$-source mixture, a $M$-output model with $M \geq N$ should output $M$ non-zero sources, and $M-N$ all-zero sources.

\subsection{Evaluation metrics}

SS systems are evaluated in terms of scale-invariant SNR (SI-SNR)~\cite{LeRoux2018a}. Since FUSS mixtures can contain one to four sources, we report two scores to summarize performance: multi-source SI-SNR improvement (MSi), which measures the separation quality of mixtures with two or more sources, and single-source SI-SNR (1S), which measures the separation model's ability to reconstruct single-source inputs.

% Combined SS+SED table for FUSS
\begin{table}[t]
\centering
\caption{SS and SED performance for FUSS-trained SS models: MSi (multi-source SI-SNR improvement) and 1S (single-source SI-SNR). Confidence intervals: $\pm$~1.2 (F1-score) and $\pm$~0.015 (PSDS).}
\begin{tabular}{|l|rrrr||c|c|}
\hline&\multicolumn{4}{c||}{FUSS test set}&\multicolumn{2}{c|}{REC\_VAL}\\
FUSS  & \multicolumn{2}{c}{Rev.} & \multicolumn{2}{c||}{Dry}&\multicolumn{2}{c|}{Late Integration}\\

 training & MSi & 1S & MSi & 1S & F1-Score & PSDS \\
  \hline
Rev. & \bf{12.5} & \bf{37.6} & \bf{10.4} & \bf{32.1}
    & 38.2&0.565\\
Dry & 10.4 & 31.2 & 10.2 & 31.8
    &\bf{39.2}&\bf{0.574}\\
\hline
\end{tabular}
\label{tab:ss_sed_fuss}
\end{table}
\begin{table}[t]
    \centering
    \caption{DESED+FUSS tasks.}
    \begin{tabular}{|l|l|}
        \hline
         Task & Sources  \\
         \hline
         DmFm & DESED mix, dry FUSS mix  \\
         BgFgFm & DESED bg, DESED fg mix, dry FUSS mix  \\
         PIT & DESED bg, dry FUSS mix, 5 DESED fg sources \\
         Classwise & DESED bg, 10 DESED classes, dry FUSS mix \\
         GroupPIT & DESED bg, 5 DESED fg sources, 4 dry FUSS srcs \\
         \hline
    \end{tabular}
    \label{tab:desedfuss}
\end{table}

% Combined SS+SED table for DESED+FUSS
\begin{table*}[t!]
\centering
\caption{SS and SED performance for various SS tasks. ``Bgd'' is DESED background, ``Fgd'' is DESED foreground, and ``Fmx'' is FUSS mixture. Confidence intervals: $\pm$~1.2 (F1-score) and $\pm$~0.015 (PSDS) on the validation set and $\pm$~2.3 (F1-score) on the synthetic set.}
\begin{tabular}{|l|rrr
%|rrr
||c|c|c|c|c|c||c|}
\hline&\multicolumn{3}{c||}{BgFgFm validation}
%&\multicolumn{3}{c||}{Dry2 eval set}
&\multicolumn{6}{c||}{REC\_VAL}&SYN\_VAL\\
SS training  & \multicolumn{3}{c||}{SI-SNRi (dB)}
%& \multicolumn{3}{c||}{SI-SNRi (dB)}
& \multicolumn{2}{c|}{Early integration}& \multicolumn{2}{c|}{Middle integration} &  \multicolumn{2}{c||}{Late integration} &Late integration \\
task  & Bg & Fg & Fm
%& Bgd & Fgd & Fmx
& F1-score & PSDS & F1-Score & PSDS & F1-Score & PSDS& F1-score \\
%   \hline
% FUSS Baseline&&&&&39.2~$\pm$~1.1&0.574~$\pm$~0.013\\
% Dry$_1$&35.4~$\pm$~0.9&0.545~$\pm$~0.012&35.9~$\pm$~1.1&0.548~$\pm$~0.012&36.2~$\pm$~1.0&0.573~$\pm$~0.014\\
% Dry$_2$&35.1~$\pm$~0.9&0.529~$\pm$~0.015&33.2~$\pm$~1.1&0.531~$\pm$~0.014&37.7~$\pm$~1.0&0.568~$\pm$~0.014\\
% Dry$_4$&31.6~$\pm$~1.0&0.461~$\pm$~0.012&33.2~$\pm$~1.2&0.472~$\pm$~0.012&37.9~$\pm$~1.1&0.574~$\pm$~0.014\\
% Dry$_4$np&27.4~$\pm$~0.9&0.361~$\pm$~0.011&28.9~$\pm$~0.9&0.386~$\pm$~0.012&38.4~$\pm$~1.1&0.566~$\pm$~0.015\\
% Dry$_6$&31.6~$\pm$~1.0&0.486~$\pm$~0.012&32.3~$\pm$~1.0&0.473~$\pm$~0.015&38.2~$\pm$~1.0&0.570~$\pm$~0.014\\
% \hline
  \hline
% FUSS& $8.1$& $13.2$& $17.9$
%     &&&&&38.2&0.565&&&\\
% DRY$_{\mathrm{FUSS}}$& $7.6$& $12.8$& ${\bf 18.0}$
%     & 32.0&0.499&&&39.2&0.574&60.8&&61.4\\
DmFm & $1.8$& $0.1$& $17.3$
    & \bf{35.4}&\bf{0.545}&\bf{35.9}&\bf{0.548}&36.2&0.573&62.3\\
BgFgFm & $\bf{18.3}$& $\bf{18.4}$& $\bf{17.5}$
    & 35.1&0.529&33.2&0.531&37.7&0.568&\bf{62.6}\\
PIT & $17.2$& $17.6$& $17.3$
    & 31.6&0.461&33.2&0.472&37.9&\bf{0.574}&62.4\\
Classwise & $16.8$& $17.5$& $\bf{17.5}$
    & 27.4&0.361&28.9&0.386&\bf{38.4}&0.566&62.0\\
GroupPIT & $16.8$& $18.0$& $17.2$
    & 31.6&0.486&32.3&0.473&38.2&0.570&62.2\\
\hline
\end{tabular}
\label{tab:ss_sed_desedfuss}
\end{table*}

\begin{figure*}
\centering
% \small
\begin{subfigure}{0.32\textwidth}
    \centering
    \includegraphics[width=0.9\linewidth]{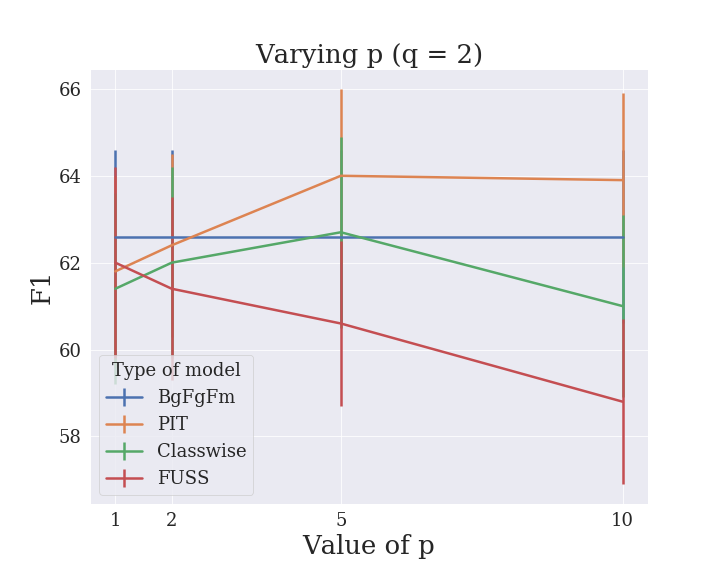}
    \caption{Impact of $p$ (SYN\_VAL)}
    \label{fig:p_synth}
\end{subfigure}
\begin{subfigure}{0.32\textwidth}
    \centering
    \includegraphics[width=0.9\linewidth]{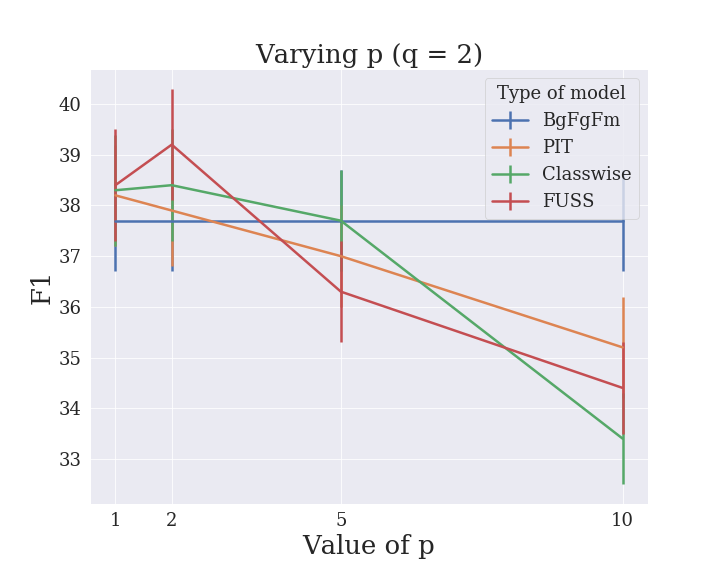}
    \caption{Impact of $p$ (REC\_VAL)}
    \label{fig:p_record}
\end{subfigure}
\begin{subfigure}{0.32\textwidth}
    \centering
    \includegraphics[width=0.9\linewidth]{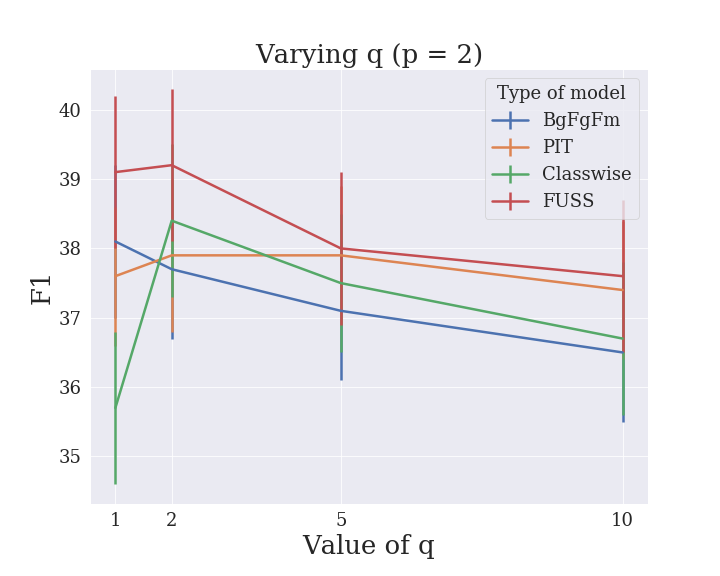}
    \caption{Impact of $q$ (REC\_VAL)}
    \label{fig:q_record}
\end{subfigure}
\caption{impact of the late integration weights on the SED performance (vertical bars represent confidence intervals)}
\label{fig:p_q_sed}
\end{figure*}

SED systems are evaluated according to an event-based F1-score with a 200~ms collar on the onsets and a collar on the offsets that is the greater of 200~ms and 20\% of the sound event's length. The overall F1-score is the unweighted average of the class-wise F1-scores (macro-average). F-scores are computed on a single operating point (decision thresholds=0.5) using the sed\_eval library~\cite{mesaros_metrics_2016}.

SED systems are also evaluated with poly-phonic sound event detection scores(PSDS)~\cite{Bilen2020}. PSDS are computed using 50 operating points (linearly distributed from 0.01 to 0.99) with the following parameters: detection tolerance parameter ($\rho_{\mathrm{DTC}}=0.5$), ground truth intersection parameter ($\rho_{\mathrm{GTC}} = 0.5$), cross-trigger tolerance parameter ($\rho_{\mathrm{CTTC}} = 0.3$), maximum False Positive rate ($e_{\mathrm{max}} = 100$). The weight on the cost trigger cost is set to $q_{\mathrm{CT}}=1$ and the weight on the class instability cost is set to $q_{\mathrm{ST}}=1$.

\section{Experiments}

Table \ref{tab:ss_sed_fuss} displays SS and SED performance on the FUSS test set and REC\_VAL. For SS we do a full cross-evaluation between the dry and reverberant versions. From this we can see that the reverberant FUSS-trained model achieves the best separation scores across both dry and reverberant conditions. However, in terms of SED performance, the dry FUSS-trained separation model yields the best performance in terms of both F1 and PSDS. This may be due to the synthetic room impulse responses used to create reverberant FUSS being mismatched to the real data in REC\_VAL. Thus, we opt to use the dry version of FUSS in the proceeding experiments.

% \subsection{Custom SS experiments}

% \begin{table}[h]
% \centering
% \caption{Per-source SS performance in SI-SNRi on Dry2 task.}
% \begin{tabular}{lrrrrrr}
% \toprule
% SS  & \multicolumn{2}{c}{Background} & \multicolumn{2}{c}{Foreground} & \multicolumn{2}{c}{FUSS mix}  \\
% % & \cmidrule(lr){2-3} & \cmidrule(lr){4-5} & \cmidrule(lr){6-7} \\
% & Mean & Std & Mean & Std & Mean & Std \\
% \midrule
% Dry1 & $1.8$ & $3.8$& $0.1$ & $6.0$& $17.3$ & $12.2$ \\
% Dry2 & $18.3$ & $5.7$& $18.4$ & $13.8$& $17.5$ & $11.9$ \\
% Dry4 & $17.2$ & $5.5$& $17.6$ & $13.5$& $17.3$ & $11.9$ \\
% Dry4np & $16.8$ & $5.6$& $17.5$ & $13.6$& $17.5$ & $11.6$ \\
% Dry6 & $16.8$ & $5.8$& $18.0$ & $13.7$& $17.2$ & $12.1$ \\
% Dry FUSS & $7.6$ & $7.2$& $12.8$ & $10.6$& $18.0$ & $11.1$ \\
% FUSS & $8.1$ & $7.0$& $13.2$ & $11.3$& $17.9$ & $11.2$ \\
% \bottomrule
% \end{tabular}
% \label{tab:ss}
% \end{table}

Besides training SS systems on FUSS, we also constructed a number of tasks consisting of data from both DESED and FUSS, described in Table \ref{tab:desedfuss}. Some tasks are trained with permutation-invariant training (PIT) \cite{yu2017permutation} or groupwise PIT.
% These tasks include DmFm with 2 sources: DESED mixture, dry FUSS mixture; BgFgFm with 3 sources: DESED background, DESED foreground mixture, dry FUSS mixture; PIT with 7 sources: DESED background, 5 DESED foreground sources, dry FUSS mixture, and trained with PIT; Classwise with 12 sources: DESED background, 10 DESED classes, dry FUSS mixture; GroupPIT with 10 sources: DESED background, 5 DESED foreground sources, 4 dry FUSS sources. Groupwise PIT is used for the groups of DESED foreground sources and FUSS sources.

Table \ref{tab:ss_sed_desedfuss} reports the results of evaluating these models on the BgFgFm task. For models with more than three outputs, the sources for corresponding classes are summed together. For example, sources 1 through 5 are summed together for the PIT and GroupPIT models, and sources 0 through 9 for the Classwise model, to produce the separated estimate of the DESED foreground mixture. The BgFgFm-trained SS model achieves the best SS scores, since it is matched to the task. This model also achieves the highest F1 score on the SYN\_VAL set, although this is not statistically significant. However, on REC\_VAL, the Classwise model achieves the best F1 score. However, notice that the dry FUSS SS model achieves the overall best F1 and PSDS scores of 39.2 and 0.574 in Table \ref{tab:ss_sed_fuss}. This suggests that the DESED+FUSS-trained SS models do not generalize as well, since they are trained on more specific synthetic data compared to FUSS-trained models.

Figure~\ref{fig:p_q_sed} displays the impact of the late integration parameters $p$ and $q$ on the SED performance. Intuitively when the SS models aims at separating sources that corresponds to target sound events, the parameter $p$ should be high so the source aggregation is close to a max pooling across sources. This is what can be observed on~Fig. \ref{fig:p_synth} for the PIT model. For the FUSS-trained SS separated sources do not correspond to target sources and the integration is better for low values of $p$. This however is not confirmed on REC\_VAL (Fig.~\ref{fig:p_record}). This could be due to the mismatch between training and test for the SS leading to sound sources that are not properly separated.

The SED performance depending on the parameter $q$ is presented on Figure~\ref{fig:q_record}. A high value for the parameter $q$ means focusing only on the mixture or on the separated sounds and leads to degraded performance for all the SS models. The best performance is then obtained with the FUSS-trained SS and $p=2$ and $q=2$ (40.7\% F1-score and 0.570 PSDS on REC\_EVAL).

\section{Conclusion}
\label{sec:conc}
In this paper we proposed to use a SS algorithm as pre-processing to a SED system applied to complex mixtures including non-target events and background noise. We proposed to retrain the generic SS on task specific datasets. The combination has shown to have potential to improve the SED performance in particular when using a late integration to combine the prediction obtained from the separated sources. However, the benefits still remain limited most probably because of the mismatch between the SS training conditions and the SED test conditions.

\section{Acknowledgements}
We would like to thank the other organizers of DCASE 2020 task 4: Daniel P. W. Ellis and Ankit Parag Shah.
% -------------------------------------------------------------------------
% Either list references using the bibliography style file IEEEtran.bst
\bibliographystyle{IEEEtran}
\bibliography{refs}
%
% or list them by yourself
% \begin{thebibliography}{9}
%
% \bibitem{dcase2016web}
%   \url{http://www.cs.tut.fi/sgn/arg/dcase2016/}.
%
% \bibitem{IEEEPDFSpec}
%   {PDF} specification for {IEEE} {X}plore$^{\textregistered}$,
%   \url{http://www.ieee.org/portal/cms_docs/pubs/confstandards/pdfs/IEEE-PDF-SpecV401.pdf}.
%
% \bibitem{PDFOpenSourceTools}
%   Creating high resolution {PDF} files for book production with
%   open source tools,
%   \url{http://www.grassbook.org/neteler/highres_pdf.html}.
%
% \bibitem{eWilliams1999}
% E. Williams, \emph{Fourier Acoustics: Sound Radiation and Nearfield Acoustic
%   Holography}. London, UK: Academic Press, 1999.
%
% \bibitem{ieeecopyright}
%   \url{http://www.ieee.org/web/publications/rights/copyrightmain.html}.
%
% \bibitem{cJones2003}
% C. Jones, A. Smith, and E. Roberts, ``A sample paper in conference
%   proceedings,'' in \emph{Proc. IEEE ICASSP}, vol. II, 2003, pp. 803--806.
%
% \bibitem{aSmith2000}
% A. Smith, C. Jones, and E. Roberts, ``A sample paper in journals,''
%   \emph{IEEE Trans. Signal Process.}, vol. 62, pp. 291--294, Jan. 2000.
%
% \end{thebibliography}

\end{sloppy}
\end{document}